\documentclass{article}
\usepackage{amsmath,graphicx,mlspconf}

\title{\Large Investigating Brain Connectivity and Regional Statistics from EEG for early stage Parkinson's Classification}
\name{%
    Amarpal Sahota, Amber Roguski, Matthew W Jones, Zahraa S. Abdallah, Raul Santos-Rodriguez
}
\address{%
    University of Bristol, UK \\
    amarpal.sahota@bristol.ac.uk, zahraa.abdallah@bristol.ac.uk, enrsr@bristol.ac.uk
}

\begin{document}

\maketitle

\begin{abstract}
We evaluate the effectiveness of combining brain connectivity metrics with signal statistics for early stage Parkinson's Disease (PD) classification using electroencephalogram data (EEG). The data is from 5 arousal states - wakeful and four sleep stages (N1, N2, N3 and REM). Our pipeline uses an Ada Boost model for classification on a challenging early stage PD classification task with with only 30 participants (11 PD , 19 Healthy Control). Evaluating 9 brain connectivity metrics we find the best connectivity metric to be different for each arousal state with Phase Lag Index achieving the highest individual classification accuracy of 86\% on N1 data. Further to this our pipeline using regional signal statistics achieves an accuracy of 78\%, using brain connectivity only achieves an accuracy of 86\% whereas combining the two achieves a best accuracy of 91\%. This best performance is achieved on N1 data using Phase Lag Index (PLI) combined with statistics derived from the frequency characteristics of the EEG signal. This model also achieves a recall of 80 \% and precision of 96\%. Furthermore we find that on data from each arousal state, combining PLI with regional signal statistics improves classification accuracy versus using signal statistics or brain connectivity alone. Thus we conclude that combining brain connectivity statistics with regional EEG statistics is optimal for classifier performance on early stage Parkinson's. Additionally, we find outperformance of N1 EEG for classification of Parkinson's and expect this could be due to disrupted N1 sleep in PD. This should be explored in future work.

\end{abstract}

\section{Introduction} \label{intro}
Parkinson's Disease is the second most prevalent neurodegenerative disease and is characterised by the loss of dopaminergic neurons in the substantia nigra \cite{han2013investigation} \cite{yuvaraj2018novel}. Extensive loss of these neurons culminates in movement-related symptoms including muscle stiffness, instability and a tremor.  However, prior to the development of motor symptoms, PD is very challenging to diagnose, complicating early detection and intervention.
The progression of the disease can be rated accorded to a clinical rating scale such as the HY (Hans and Yahr) scale \cite{bhidayasiri2012parkinson}. This scale defines broad categories of motor function in Parkinson's and has 5 stages with stage 1 being the earliest. There is a need for earlier diagnosis of Parkinson's and Electroencephalogram (EEG)  has shown promise for this. EEG is non-invasive, relatively inexpensive and records the electrical activity of cortical neurons in the brain with a high temporal resolution.

Limited work has been done on the earliest stage (HY 1) Parkinson's classification. Limited work has also been done on the fusion of brain connectivity with signal statistics and exploration on whether the information in each is supplementary or redundant. Prior work on general PD classification uses methods such as deep learning (e.g. Convolutional Neural Networks), signal statistics input to a machine learning model and measures of brain connectivity (see related work \ref{related work}) input to a machine learning model. Machine learning models used in this previous work include Support Vector Machines, Neural Networks and tree based models such as Random Forest and Ada Boost classifiers \cite{sahota2023interpretable}.

In this work we evaluate the effectiveness of combining signal statistics with brain connectivity for early stage PD classification (mean HY 1.55) using EEG recorded across 5 arousal states. The arousal states include wakeful data and the four sleep stages (N1, N2, N3 and REM). We investigate 9 brain connectivity metrics individually and in combination with regional EEG statistics. Best classifier performance is achieved by combining brain connectivity with signal statistics from multiple brain regions. This result suggests that brain connectivity metrics and signal statistics provide supplementary information for classification of early stage PD and best performance is achieved by combining both.

\section{Related Work} \label{related work}

\subsection{Parkinson's Classification with EEG}
Prior work has found frequency based statistics of the EEG signal input to machine learning models to be effective for classification. We build on work regarding peak frequencies and band powers from EEG. Band powers refers to EEG signal strength within defined frequency ranges whereas peak frequency refers to the highest strength frequency \cite{sahota2023interpretable}.  In \cite{chaturvedi2017quantitative} the authors achieve 78\% accuracy on PD classification with a balanced data set using EEG band power values, median and peak frequency as features. In  \cite{betrouni2019electroencephalography} an accuracy of 86\% is achieved with a support vector machine using frequency based EEG features. In  \cite{waninger2020neurophysiological} the authors combine coherence (functional connectivity measure) with wavelet coefficients achieving an accuracy of 94\%. More complex spectral features have also been used such as in \cite{yuvaraj2018novel} where authors achieve 99\% accuracy using features of the third-order spectra of EEG (bispectrum). However, a limitation of this study is that participants were not in the early stage of PD - they were in stage II and III according to the HY scale. Other work such as  \cite{khoshnevis2021classification} achieved lower performance (85\% accuracy) using complex spectral features. Deep Learning methods have also shown promise such as in \cite{lee2019deep} where the authors achieve 96.9\% accuracy using a 'convolutional-recurrent' neural network architecture. To summarise, EEG classification of Parkinson's shows promise. However current work on early stage Parkinson's is limited and limited work has been done to explore different connectivity metrics across different arousal states and how performance changes when fusing brain connectivity information with regional statistical EEG features.  

\subsection{Brain Connectivity} \label{brain_connect_sec}
Measures of connectivity across the brain provide valuable information about brain activity and are thus useful for EEG classification. Different types of connectivity include structural connectivity, functional connectivity and effective connectivity. Structural connectivity corresponds to anatomical connectivity between brain regions, functional connectivity characterises statistical dependencies or correlations in activity and effective connectivity explores causal or directional influences between regions\cite{chiarion2023connectivity}. A number of brain connectivity measures have been defined in the EEG literature \cite{chiarion2023connectivity}. For this work we explore 9 brain connectivity measures including  Coherence (coh) , Imaginary Coherence (imcoh) , Phase-Locking Value (PLV), corrected imaginary PLV (ciplv) , Pairwise Phase Consistency (PPC) , Phase Lag Index (PLI), Directed Phase Lag Index (DPLI), Weighted Phase Lag Index (WPLI) and debiased weighted phase lag index (DWPLI).  

For conciseness we define PLI in detail only as this metric was evaluated in most detail in our experiments (see section \ref{Experiment}). Phase lag index measures the asymmetry in the distribution of phase differences over time between two signals. The PLI ranges between 0 and 1 with 0 indicating no coupling or instaneous coupling and 1 indicating that there is a true lagged interaction. By definition if the distribution of phase differences between two signals are given by $\Delta \phi(t_k)$ for $k = 1 \ldots N$ then the phase lag index is given by \cite{kuang2022phase} : $\text{PLI} = \left| \left\langle \text{sign}\left[\Delta \phi(t_k)\right] \right\rangle \right|]$ 

Connectivity metrics such as PLI can be calculated for two signals or for components of each signal. For example, one can calculate the PLI between the delta band (0.5-4Hz) of one signal with the delta band of another signal. In \cite{kuang2022phase} the authors use Phase Lag Index to classify Mild cognitive impairment and find phase lag index in the alpha band to be correlated with a cognitive assessment score.

PLI as a connectivity metric has the benefit of being insensitive to volume conduction effects \cite{cohen2015effects}. Volume conduction occurs when there is a single source of neuronal activity in the brain but it's electrical signal is measured simultaneously at multiple electrodes. This can lead to a false positive for connectivity between electrodes or brain regions.

\subsection{Frequency based signal statistics} \label{freq stats}
Frequency based statistics are commonly used in EEG classification pipelines. Power spectral density (PSD) refers to the distribution of power vs. frequency for a signal and is computed with the Fourier Transform. Research has shown the strength of the EEG signal in specific frequency power bands to be informative of brain activity. These power bands are defined as delta (0.5--4Hz), theta (4--8Hz), alpha (8--12Hz), sigma (12--16Hz), beta (16--30Hz) and gamma (30--40Hz) \cite{bandpowerpaper}. The peak frequency (highest power frequency) in the PSD aswell as the power in that frequency are also informative statistics for classification. For longer EEG signals (over 5 minutes) enhanced classification performance can be achieved by calculating distributions of these frequency based statistics over time windows across the signal. \cite{sahota2023interpretable}

\section{Method}
\label{method}

The data used in this work is from a yet unpublished dataset containing 57 channel wakeful and sleep EEG from Parkinson's and Healthy participants. The data set is described in further detail in \ref{Data}.

\subsection{Preprocessing}
The EEG data for each participant was scored manually in 30-second epochs in line with AASM guidelines \cite{berry2017aasm}.  Epochs were labelled as Wake, N1, N2, N3 or REM. The EEG data was downsampled from the recording frequency of 512Hz to 256Hz and a 0.25 - 40Hz bandpass filter applied. Next, Independent Component Analysis and manual inspection of the EEG signal and power spectral density was completed for further artefact removal. Bad channels were interpolated and the data re-referenced using the REST referencing technique \cite{dong2017matlab,yao2001method}.

\subsection{Regional Statistics Classifier}  \label{stat_pipeline}
\begin{figure}[h!]
        \centering
        \includegraphics[width=0.9\columnwidth]{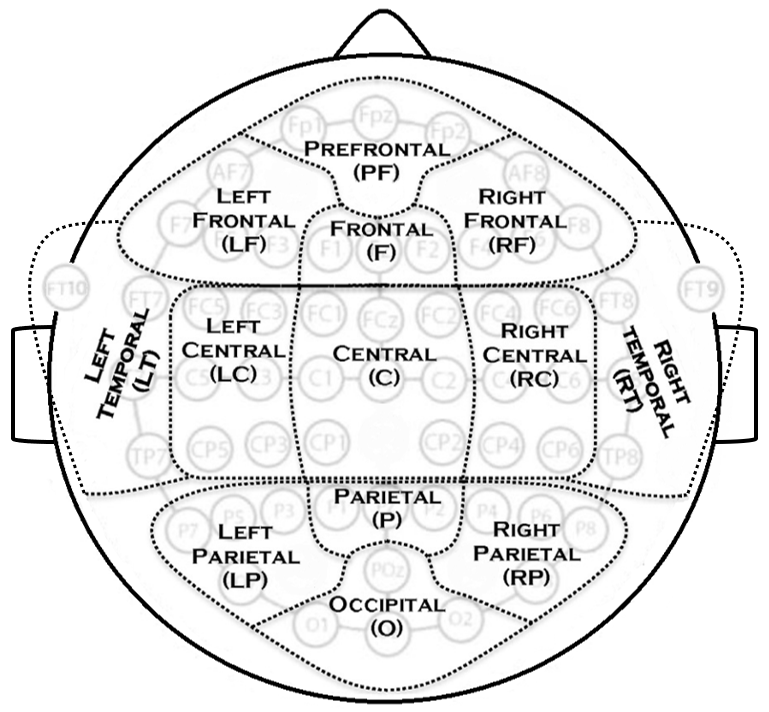}
        \caption{EEG channels grouped into 13 brain regions of interest \cite{regionaleeg} \cite{sahota2023interpretable}.}
        
        \label{brain_regions_fig}
\end{figure}

For the regional classification pipeline the 57 EEG electrodes were grouped into 13 brain regions of interest (see figure \ref{brain_regions_fig}) . 

\begin{figure}[h!]
        \centering
        \includegraphics[width=1\columnwidth]{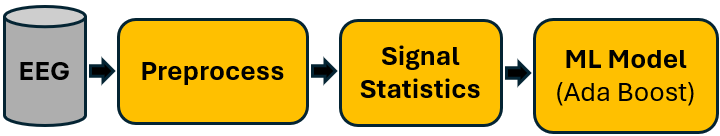}
        \caption{Pipeline for signal statistics classifier.}
        
        \label{stats_pipeline_fig}
\end{figure}

Per electrode and per region, statistics of 15 key frequency based features were calculated. These 15 statistics were the 6 band powers as defined in \ref{freq stats}, total signal power aswell as the four peak frequencies and their corresponding power values. These 15 statistics are calculated per 6 second window resulting in 15 feature distributions. Key statistical features of these feature series including length, maximum, absolute maximum, minimum, mean, median, standard deviation, variance, root mean square and sum are calculated and input to a machine learning model (Ada Boost). The pipeline is shown in figure \ref{stats_pipeline_fig}.

The statistical model is fit using features from single brain regions, two brain regions together at a time and all brain regions together.

\subsection{Connectivity Classifier} \label{connect_pipeline}
\begin{figure}[h!]
        \centering
        \includegraphics[width=1\columnwidth]{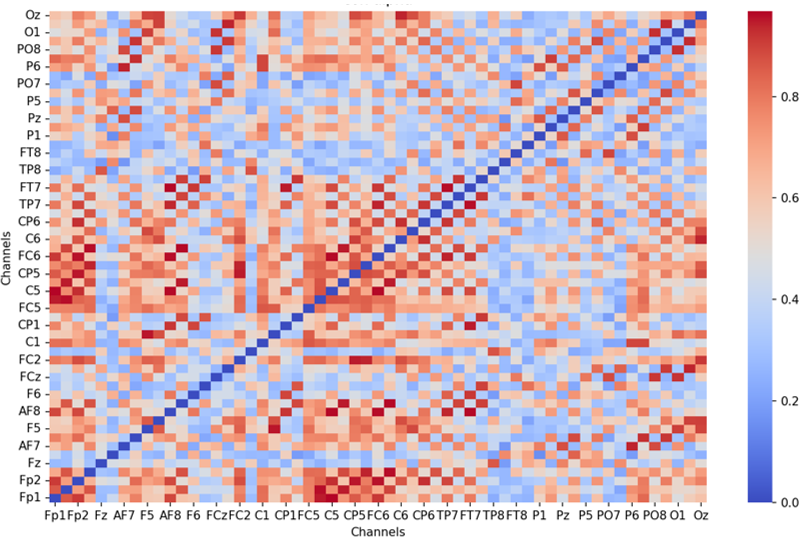}
        \caption{Example connectivity grid with electrodes (channels) along the x and y axes. Grid is filled with connectivity value for each electrode-electrode pair. Colour gradient fill with blue as 0 and dark red as 1.}
        
        \label{connectiviy_grid}
\end{figure}

For the connectivity classification pipeline. Connectivity is calculated for every electrode-electrode pair. For 57 electrodes, this results in 1596 features calculated per sample. An example electrode-electrode connectivity grid is shown in Figure \ref{connectiviy_grid}. The x-axis and y-axis contain the electrode names and between each electrode a connectivity value is calculated. The grid is symmetrical as electrode 1 connectivity with electrode 2 is the same as electrode 2 connectivity with electrode 1. The diagonal is blank as the connectivity of an electrode is not calculated with itself.

The calculated connectivity features are input to a machine learning model (AdaBoost) for classification. This is done for 9 connectivity metrics as outlined in experimental procedure (section \ref{Experiment}).

\begin{figure}[h!]
        \centering
        \includegraphics[width=1\columnwidth]{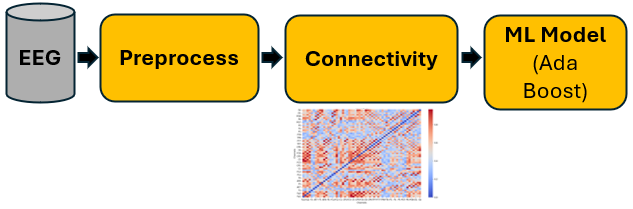}
        \caption{Pipeline for connectivity classifier.}
        
        \label{connectivity_pipeline_fig}
\end{figure}

\subsection{Combined Classifier}
The combined classifier simply combines the regional statistical features with brain connectivity features for classification. Both sets of features are input to our machine learning model (Ada Boost classifier).

\section{Experiments and Discussion}
\label{results}

\subsection{Data} \label{Data}
The dataset includes 57-channel EEG data for 19 Healthy Control and 11 Parkinson's participants.  The data includes wakefulness EEG and recordings of the four sleep stages (N1, N2, N3, REM). In home recording was conducted on 2 or 3 nights for all participants in line with the AASM standards for polysomnography  \cite{berry2017aasm}. The demographics of PD and Healthy participants are similar, see table \ref{data_stats}

\begin{table}[h!]
\centering
\begin{tabular}{ |c|c|c|c| } 
  \hline
 - & Healthy Control (19) &  Parkinson's (11)  \\
 \hline
 Female & 5 (26.3\%) &  4 (36.4\%)   \\
 Male & 14 (73.7\%) &  7 (63.6\%)  \\
 Age & 69.57 $\pm$ 8.77 & 67.82 $\pm$ 10.77  \\ 
 HY Stage & - &  1.55 $\pm$ 0.21  \\ 
\hline
\end{tabular}
\caption{Age and sex of participants aswell as mean HY stage for Parkinsons patients.}
\label{data_stats}
\end{table}

The number of samples exceeds the number of participants as EEG data was recorded on multiple nights. Those recordings of a low quality were dropped. Samples per class per arousal state aswell as mean sample length in epochs (where an epoch is 30s for sleep data and 20s for wakeful) are shown in Table  \ref{data_stats_2}.

\begin{table}[h!]
\begin{center}
\begin{tabular}{ |c|c|c| } 
  \hline
 Arousal state  &  Samples & Length (epochs)  \\
 \hline
 Wakeful  &  HC 17 , PD 11 & 10 $\pm$ 2  \\
 N1   &  HC 26 , PD 15 & 30 $\pm$ 23 \\
 N2  &  HC 26 , PD 15 & 300 $\pm$ 110\\ 
 N3  &  HC 26 , PD 15 & 150 $\pm$ 54\\ 
 REM   &  HC 26 , PD 14 & 105 $\pm$ 40 \\ 
\hline
\end{tabular}
\caption{ Samples per class and mean length of recording in epochs per arousal state.}
\label{data_stats_2}
\end{center}
\end{table}

\subsection{Experimental Procedure} \label{Experiment}
All experiments were run with two random seeds using 5-fold cross validation. The mean across the two sets of 5-fold cross validated experiments was calculated to determine the classifier performance. Where multiple samples were present from a single participant they were kept in the same fold to avoid information leakage. 

The regional statistics classification pipeline (see \ref{stat_pipeline}) was deployed using features from each of the 13 brain regions individually, using features from two brain regions at a time (78 combinations) and using features from all 13 brain regions at once for each arousal state. 

The connectivity pipeline (see \ref{connect_pipeline} ) was deployed for 9 connectivity metrics for each arousal state. Each connectivity metric was calculated for each band power delta (0.5–4Hz), theta (4–8Hz), alpha (8–12Hz), sigma (12–16Hz), beta (16–30Hz), gamma (30–40Hz) and for the entire EEG signal. Thus there were 7 results for each connectivity metric for each arousal state. The 9 metrics used in the pipeline were Coherence (CoH), Imaginary Coherence (Imcoh), Phase-Locking Value (PLV), corrected imaginary PLV, Pairwise Phase Consistency (PPC) , Phase Lag Index (PLI), Unbiased Phase Lag Index, Directed Phase Lag Index (DPLI), Weighted Phase Lag Index (WPLI) and debiased weighted phase lag index (DWPLI).  

Finally, a combined pipeline was deployed for the best performing overall connectivity metric combined with each set of statistical features for each arousal state.

\subsection{Results} \label{Results}

The regional statistical features pipeline had best performance using features from two brain regions at a time. We present results from this best performing statistical approach. For this classification method a best accuracy of 78.3\% was achieved on Wakeful data and 77.5 \% accuracy on  N1 data. Worse performances (less than 65 \% accuracy) were achieved on N2, N3 and REM data (see column 2 in table \ref{combined_results}).

The best performing connectivity metric was Phase Lag Index on the gamma power band using N1 data which achieved a performance of 86.2\% classification accuracy. Across other arousal states other metrics performed better. Table \ref{best_con_metrics} below shows the best performing metric per arousal state with their corresponding performance.

\begin{table}[h!]
\centering
{\footnotesize 
\setlength{\tabcolsep}{4pt} 
\begin{tabular}{|c|c|c|c|c|c|} 
\hline
- & Wake & N1 & N2 & N3 & REM \\
\hline
Metric & CoH & PLI & WPLI & Imcoh & DWPLI \\
Acc. (\%) & 68.3$\pm$15.9 & 86.2$\pm$10.4 & 73.3$\pm$15 & 73.8$\pm$16.8 & 68.8$\pm$11.6 \\
\hline
\end{tabular}
}
\caption{ Best performing connectivity metric per arousal state and accuracy. Acronyms for metrics are defined in \ref{brain_connect_sec}.}
\label{best_con_metrics}
\end{table}

With PLI as the highest performing connectivity metric on an individual arousal state (86.2 \% N1) we combined PLI with statistical features across arousal states. Results are in table \ref{combined_results} below. The table also shows results for the best individual PLI only models and best statistical features only models to highlight the effect of combination.

\begin{table}[h!]
    \centering
    \begin{tabular}{|p{0.2\linewidth}  |p{0.2\linewidth}  |p{0.2\linewidth} |p{0.2\linewidth} |} \hline

      Arousal state  & Regional Features & Connectivity (PLI) & Regional Feats \& PLI  \\ \hline  
      Wake &  78.3 $\pm$16.9 &  56.7 $\pm$17.1  &   80 $\pm$19.6\\ \hline 
      N1 &  77.5 $\pm$11.0 &  86.2 $\pm$10.4  &    \textbf{91.3 $\pm$8.1} \\ \hline 
      N2 &  57.9 $\pm$18.7 &  66.9 $\pm$11.4  &   73.3 $\pm$12.9\\ \hline 
      N3 &  60.7 $\pm$22.4  &  64.7 $\pm$16.5  &   69.3 $\pm$20.5\\ \hline 
      REM &  63.7 $\pm$19.0  &  63.8 $\pm$13.1  &   66.2 $\pm$14.9\\ \hline

    \end{tabular}
    \caption{Model accuracy (\%) per arousal state when using regional statistical features only, connectivity (PLI) only and then combining PLI and regional statistical features.}
    \label{combined_results}
\end{table}

\subsubsection{Analysis}
Per table \ref{best_con_metrics} the best performing connectivity metric varied per arousal state. The best overall performance was with N1 data where Phase Lag Index on the gamma power band achieved an accuracy of 86.2\%. Phase Lag Index measures the consistency of phase differences between two signals and is insensitive to volume conductive effects (see \ref{brain_connect_sec}). Other metrics derived from PLI include Weighted Phase Lag Index which de-weights the value of small phase differences and the debiased estimator of squared weighted phase lag index (DWPLI). With PLI having highest performance for N1 this suggests that volume conduction effects are likely to be present in the data as well as true lagged interaction at small phase differences. This would explain why WPLI and the debiased estimator of squared WPLI do not outperform PLI on N1 data. Across other arousal states different metrics perform best. This suggests that the patterns of connectivity discriminating Healthy Controls from those with Parkinson's likely varies across arousal states. This should be explored in further detail for future work. 


Table \ref{combined_results} shows per arousal state the best accuracy achieved with using Statistical Features only, Connectivity (PLI) only and Statistical Features combined with Phase Lag Index. For every arousal state combining statistical features with PLI leads to an increase in mean accuracy. This suggests that the statistical features and PLI provide useful complementary information for classification of Parkinson's. Therefore if performance is the main priority over computational efficiency then one should always include connectivity data to complement statistical features for classification. 

The overall best performing classifier on N1 data used Statistical Features combined with Phase Lag Index. This model achieved an accuracy of 91 \%. For the classification of Parkinson's disease precision and recall are also important. This model achieved a recall of 80 \% and precision of 96\% (see figure \ref{connectiviy_grid} for confusion matrix).  

This demonstrates the effectiveness in combining connectivity and statistical features. This also indicates N1 data as being very discriminatory for this classification task. This could be due to the Parkinson's disease participants having more disrupted light sleep during the N1 phase and should be explored in future work.

\begin{figure}[h!]
        \centering
        \includegraphics[width=0.81\columnwidth]{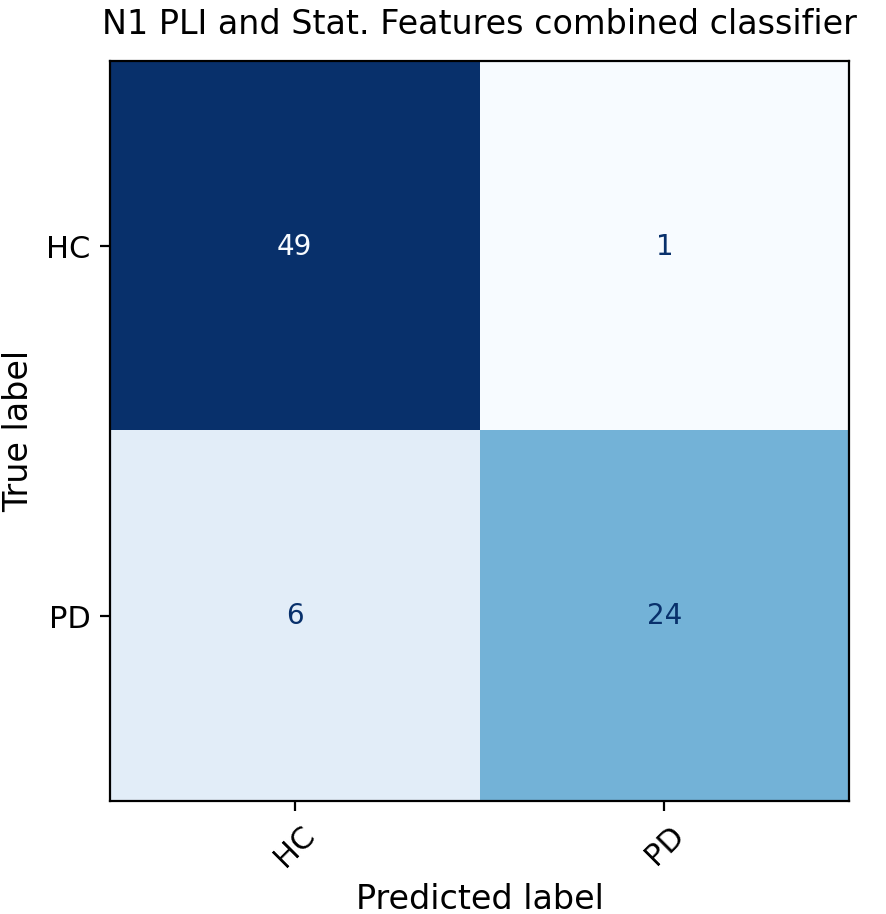}
        \caption{Confusion Matrix for highest accuracy model (91.3\%). Model uses PLI features and regional statistical features from N1 data. Model run with two random seeds, therefore samples in matrix are double the number of total samples. }
        
        \label{CM_best_model}
\end{figure}



\subsection{Future Work}
Future work should explore the differences between connectivity metric performance across data types for Parkinson's classification. This could help to highlight the differences in the Parkinson's brain activity vs Healthy controls across wakefulness and the sleep stages. 

Secondly, future work should work on optimal methods for fusion between connectivity metrics and statistical features. In this work we used simple feature concatenation (feature level fusion) to demonstrate the effect of using the respective features individually and together. For optimal classifier performance data fusion techniques such as late fusion (fusion at the decision level), model level fusion (at the representation level) or a hybrid between methods can be implemented \cite{pawlowski2023effective}.

Finally, work should be expanded to other data sets as our data set is relatively small with 30 participants (19 Healthy Control, 11 Parkinson's). Replication of our results should be done on additional data sets to show generalisability of the result.

\section{Conclusion} \label{Conclusion}

We evaluate the effectiveness of EEG signal statistics and brain connectivity metrics across 5 arousal states (Wakeful, N1, N2, N3 and REM) for early stage Parkinson's classification. Brain connectivity metrics are evaluated on their own and in combination with regional signal statistics using feature level fusion. Phase Lag Index on the gamma band power on N1 data was found to be the highest performing connectivity metric (86 \% accuracy) with different metrics performing best on each arousal state. Phase Lag Index was then combined with signal statistics across all arousal states. The combination of PLI with regional signal statistics increased mean accuracy of the classifier across every arousal state. This suggests that fusion of connectivity with regional signal statistics is important for best performance. The best overall classifier (PLI and regional statistics combined) was on N1 data achieving 91 \% accuracy, 96\% precision and 80\% recall. Future work should explore how connectivity metric performance varies across arousal state and optimal methods for fusion of EEG signal statistics with brain connectivity information.

\bibliographystyle{IEEEbib}
\bibliography{refs}

\end{document}